\documentclass{moriond}

\def\Journal#1#2#3#4{{#1} {\bf #2}, #3 (#4)}

\def\PLB{{\em Phys. Lett.}  B}

\def\PRD{{\em Phys. Rev.} D}

\def\mco{\multicolumn}

\def\ra{\rightarrow}

\def\ko{K^0}

\def\be{\begin{equation}}
\def\ee{\end{equation}}
\def\bea{\begin{eqnarray}}
\def\eea{\end{eqnarray}}

\newcommand{\Photo}

\begin{document}
\vspace*{4cm}
\title{Anatomy of a new anomaly in non-leptonic B decays}

\author{ M. Alguer\'o$^a$, A. Crivellin$^{(b,c,d)}$, S. Descotes-Genon$^e$, {\bf J. Matias}$^*\,$\footnote{*speaker}, M. Novoa-Brunet$^e$}

\address{(a) Universitat Aut\`onoma de Barcelona and IFAE, 08193 Bellaterra, Barcelona,\\
(b) CERN Theory Division, CH--1211 Geneva 23, Switzerland, 
(c) Physik-Institut, Universit\"at Z\"urich, Winterthurerstrasse 190, CH--8057 Z\"urich, Switzerland,
(d) Paul Scherrer Institut, CH--5232 Villigen PSI, Switzerland,
(e) Universit\'e Paris-Saclay, CNRS/IN2P3, IJCLab, 91405 Orsay, France \medskip}

\maketitle\abstracts{We discuss the anatomy of the $L_{VV}$ observable designed as a ratio of the longitudinal components of $B_s \to VV$ versus $B_d \to VV$ decays. We focus on the particular case of $B_{d,s} \to K^{*0} {\bar K^{*0}}$ where we find for the SM prediction $L_{K^* \bar{K}^*}=19.5^{+9.3}_{-6.8}$ implying a 2.6$\sigma$ tension with respect to data. The interpretation of this tension in a model independent way identifies two Wilson coefficients  ${\cal C}_{4}$ and ${\cal C}_{8g}$ as possible sources. The example of one simplified model including a Kaluza-Klein (KK) gluon is discussed. This KK gluon embedded inside a composite/extra dimensional model combined with a $Z^\prime$ can explain also the $b\to s\ell\ell$ anomalies albeit with a significant amount of fine-tuning.
}

\section{Motivation}
Besides the flavour anomalies in semileptonic B decays, where different New Physics (NP) hypotheses exhibit pulls above 7$\sigma$ w.r.t. the SM~\cite{global-fit}, it is natural to expect other possible signals of NP in other observables governed by the $b \to s$ transition. A natural place to look at are non-leptonic B decays. However, these decays suffer from large uncertainties compared to semileptonic B decays and traditional observables like branching ratios or polarization fractions are very difficult to predict with accuracy. 
In Ref.~\cite{non-leptonic} looking for cleaner observables, we draw a parallel between Lepton Flavour Universality Violating (LFUV) ratios  and U-spin ratios of optimized observables in non-leptonic B decays constructed upon longitudinal amplitudes of $B \to VV$ decays. While the former consists of ratios of decays to 2nd generation  leptons  ($b \to s \mu^+\mu^-$) versus 1st generation ($b \to s e^+e^-$), the latter are 
 U-spin ratios to 2nd generation quarks ($b \to s$) versus 1st generation ($b \to d$). This parallelism has two evident limitations: a) the breaking of LFUV can be more accurately computed   than the breaking of U-spin that requires hadronic contributions to be estimated and b) the remanent long distance sensitivity to weak annihilation (WA) and infrared divergent hard spectator scattering (HSS) in the non-leptonic case. 
 Our goal in Ref.~\cite{non-leptonic} was to build cleaner observables for non-leptonic modes exhibiting deviations with respect to the SM, reconsider the SM estimates and discuss potential NP sources.

\section{Theoretical Framework and hadronic uncertainties}
The theoretical description of $B_Q \to VV$ with $Q=d,s$ follows NLO QCD-Factorization (QCDF)~\cite{yang} including the modeling of the $1/m_b$ suppressed IR divergences coming from WA and HSS. 
We start by decomposing the $\bar B_Q$ decay amplitude through a $b \to q$  transition into a $V_1V_2$ state  with the same definite polarisation for both vector mesons: 
\begin{equation}
\bar{A}_f\equiv A(\bar{B}_Q\to V_1 V_2)
  =\lambda_u^{(q)} T_q + \lambda_c^{(q)} P_q\,.
\label{dec}
\end{equation}
There are three possible helicity amplitude states for the outgoing $V V$ pair.
The naive hierarchy~\cite{Kagan} among the helicity amplitudes shows that transverse amplitudes are power suppressed w.r.t the longitudinal (the electromagnetic effects known to violate this hierarchy~\cite{yang} have no impact on the following argument). The central point here is that the dangerous  IR divergences enter at leading order in the transverse amplitudes but are power-suppressed for the longitudinal one. This suggested the idea of constructing an observable only sensitive to the longitudinal amplitudes. 
We then decided to focus on decays purely mediated through penguin diagrams (i.e. no tree contributions). Indeed the theoretical uncertainties on the hadronic contributions can be reduced in these cases due to the fact that the difference of hadronic $\Delta_q=T_q-P_q$~\cite{old,delta} is free from $1/m_b$-suppressed long-distance divergences.

\subsection{The L-observable: SM prediction and comparison with data}
We define an observable that will be sensitive to the information on the polarization fraction but with a cleaner theoretical prediction~\cite{non-leptonic}: 
\begin{equation}\label{eq:LKstKst0}
L_{V_1V_2}=\frac{{\cal B}_{b \to s}}{{\cal B}_{b \to d}}\frac{g_{b \to d} f_L^{b \to s}}{g_{b \to s} f_L^{b \to d}}=\frac{|A_0^s|^2+ |\bar A_0^s|^2}{|A_0^d|^2+ |\bar A_0^d|^2}\,,
\end{equation}
where 
${\cal B}_{b \to q}$ ($f_L^{b \to q}$) refers to the branching ratio (longitudinal polarisation) of the $\bar{B}_Q \to V_1V_2$ decay governed by a $b \to q$ transition. $\bar{A}_0^q$ and ${A}_0^q$ are the 
amplitudes for the $\bar{B}_Q$ and ${B}_Q$ decays governed by $b \to q$ with final vector mesons longitudinally polarised and $g_{b \to q}$ stands for the phase space factor (see Ref.~\cite{non-leptonic} for definition). 
In the particular case of the penguin-mediated decays
 $B_{d,s} \to K^{*0} {\bar K^{*0}}$ taking experimental data from LHCb~\cite{exp1} and Babar~\cite{exp2} we find:
\begin{equation}\label{eq:expL}
   {\rm Exp}:\qquad L_{K^*\bar{K}^*}=4.43\pm 0.92,
\end{equation}
where we included 
the effect of $B_s$ meson mixing in the measurement of the branching ratio (leading to a correction of at most 7$\%$).
Concerning the SM prediction, we found in Ref.~\cite{non-leptonic}:
  \begin{equation}\label{eq:LKstKstDeltaP}
\!\! L_{K^*\bar{K}^*}=  \kappa \left|\frac{P_s}{P_d}\right|^2 
 \left[\frac{1+\left|\alpha^s\right|^2\left|\frac{\Delta_s}{P_s}\right|^2
 + 2 {\rm Re} \left( \frac{ \Delta_s}{P_s}\right) {\rm Re}(\alpha^s) 
 }{1+\left|\alpha^d\right|^2\left|\frac{\Delta_d}{P_d}\right|^2
  + 2 {\rm Re} \left( \frac{ \Delta_d}{P_d}\right) {\rm Re}(\alpha^d)} \right]\,.
 \end{equation}
 See Ref.~\cite{non-leptonic} for explicit definitions of $\kappa$ and $\alpha^{d,s}$.
We obtain in the NLO-QCDF case \cite{non-leptonic}:
 \begin{eqnarray}\label{eq:tension1}
  {\rm QCD\ fact}:    L_{K^*\bar{K}^*}=& 19.5^{+9.3}_{-6.8} \qquad 2.6 \sigma\,. 
  \label{eq:tension3}
  \end{eqnarray}
 In Table\ref{tab:confidence} the 1$\sigma$ and 2$\sigma$ confidence intervals are provided using the whole distribution for $L$.
 The $1/m_b$ suppressed contributions entering $L_{K^*\bar{K}^*}$ are parametrised in the same manner as in Ref.~\cite{yang}, involving two regulators $X_H$ and $X_A$ treated as universal
 for all channels:
  \begin{equation}
     X_{H,A}=(1+\rho_{H,A}e^{i\varphi_{H,A}})\ln{\left(\frac{m_B}{\Lambda_{h}}\right)}\,.
 \end{equation}
 The comparison between theory and experiment points to a possible deficit in the $b \to s$ transition 
 (reminiscent of the deficit for muons in $b\to s\ell\ell$ decays). Finally, a detailed analysis of the error budget points  
 to the amount of SU(3) breaking in the form factors  as the main source of uncertainty in the ratio $L_{K^*\bar{K}^*}$ (around 30$\%$) and a much lower impact of IR divergences in the ratio than in the individual penguin amplitudes. This means that improving on form factors and their correlation can substantially help in reducing the theory uncertainty  on the SM prediction. 
 \begin{table}[h]
	\begin{center}
\begin{tabular}{|c|c|c|}
\hline
Observable   & 1$\sigma$ & 2$\sigma$  \\
\hline
$L_{K^*\bar{K}^*}$ & $[12.7, 28.8]$ & $[7.5, 43]$ \\
\hline
\end{tabular}
	\caption{1$\sigma$ and 2$\sigma$ confidence intervals for the SM prediction of $L_{K^*\bar{K}^*}$ within QCD factorisation.}
		\label{tab:confidence}
    \end{center}
\end{table}

\section{Model independent interpretation and simplified models}
A model independent analysis 
 in terms of NP contributions to the Wilson coefficients of the operators of the ${\cal H}^{\rm eff}$ (governing the $b \to s$ transition~\cite{non-leptonic}) identified three possible relevant coefficients: a) the  coefficient ${\cal C}_{1s}^{ c}$ of the operator $Q_{1s}^p = (\bar p b)_{V-A} (\bar s p)_{V-A}$ 
 that however requires too large a NP contribution ($\sim$60$\%$) in conflict with recent analyses on non-leptonic constraints~\cite{Lenz:2019lvd} and dijet angular distribution bounds~\cite{dijet} b) the penguin coefficient ${\cal C}_{4s}$ of the operator $Q_{4s} = (\bar s_i b_j)_{V-A} \sum_q\,(\bar q_j q_i)_{V-A}$  that requires a NP contribution of order 25\% (incidentally of similar size as the semileptonic NP contribution to $C_9^{\rm eff}$) and c) the chromomagnetic coefficient  ${\cal C}_{8gs}^{\rm eff}$ of the operator $Q_{8gs} = \frac{-g_s}{8\pi^2}\,m_b\, \bar s\sigma_{\mu\nu}(1+\gamma_5) G^{\mu\nu} b$
 that would require a contribution of the same order of the SM, also allowed due to the very weak constraints on this coefficient.
 
 Concerning specific models to explain this tension,
 the possibility of a tree-level NP contribution entering 
  $C_{4s}$ via a massive $SU(3)_c$ Kaluza-Klein gluon (axi-gluon) was discussed. 
  We parametrise its couplings to down quarks  as
 \begin{equation} \label{lagrangian}
 {\cal L} = \Delta _{sb}^L\bar s{\gamma ^\mu }{P_L}{T^a}bG_\mu ^a + \Delta _{sb}^R\bar s{\gamma ^\mu }{P_R}{T^a}bG_\mu^a\,,
 \end{equation}
 with $\Delta_{sb}^{L,R}$ assumed real.  We also define from Eq. (\ref{lagrangian}) analogous flavour diagonal couplings which we will denote as $\Delta_{qq}^{L,R}$. We assume these universal flavour-diagonal couplings for the KK gluon to the first two generations of quarks to avoid large effects in K and D mixing. Taking maximal coupling for $\Delta^L_{qq}$ (with R partner to zero) still a significant fine-tuning is required between  $\Delta^L_{sb}$ vs $\Delta^R_{sb}$ to account for the constraint from B-meson mixing. 
 An alternative possibility through a loop-generated contribution to $C_{8gs}$ is discussed in Ref.~\cite{non-leptonic}. 

 Finally a possible model-dependent way to establish a link with $b \to s \ell\ell$ anomalies in the former model is to embed the KK gluon as part of the particle spectrum of a composite/extra-dimensional model including a $Z^\prime$ boson. While the KK gluon would explain the $L_{K^*K^*}$ tension the $Z^\prime$ due to the large $sb$ coupling could explain the $b\to s\ell^+\ell^-$ anomalies without violating LHC di-lepton bounds~\cite{Aad:2020otl}. 
 
\begin{figure}[b]
\includegraphics[width=0.5\textwidth]{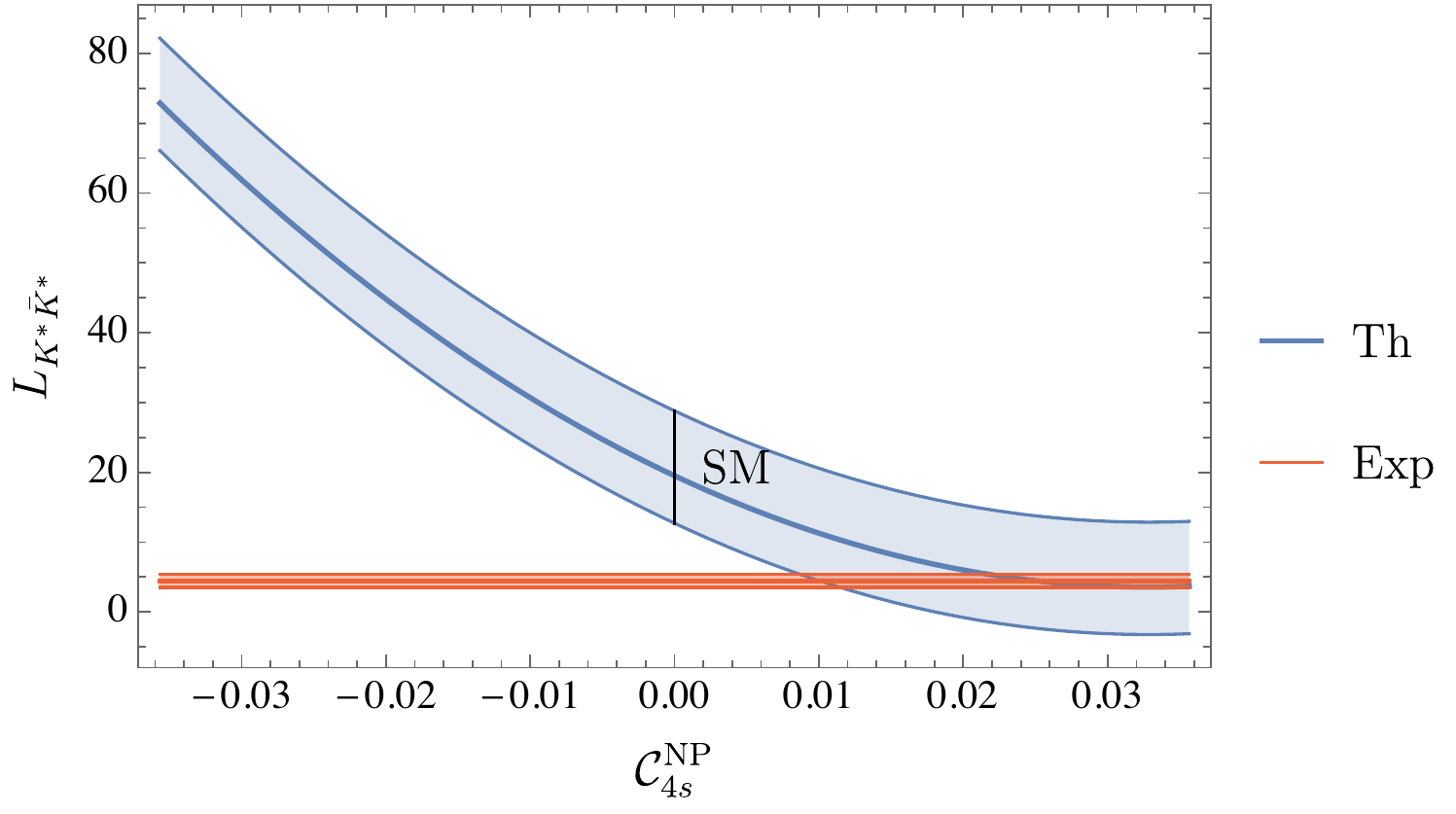}
\includegraphics[width=0.5\textwidth]{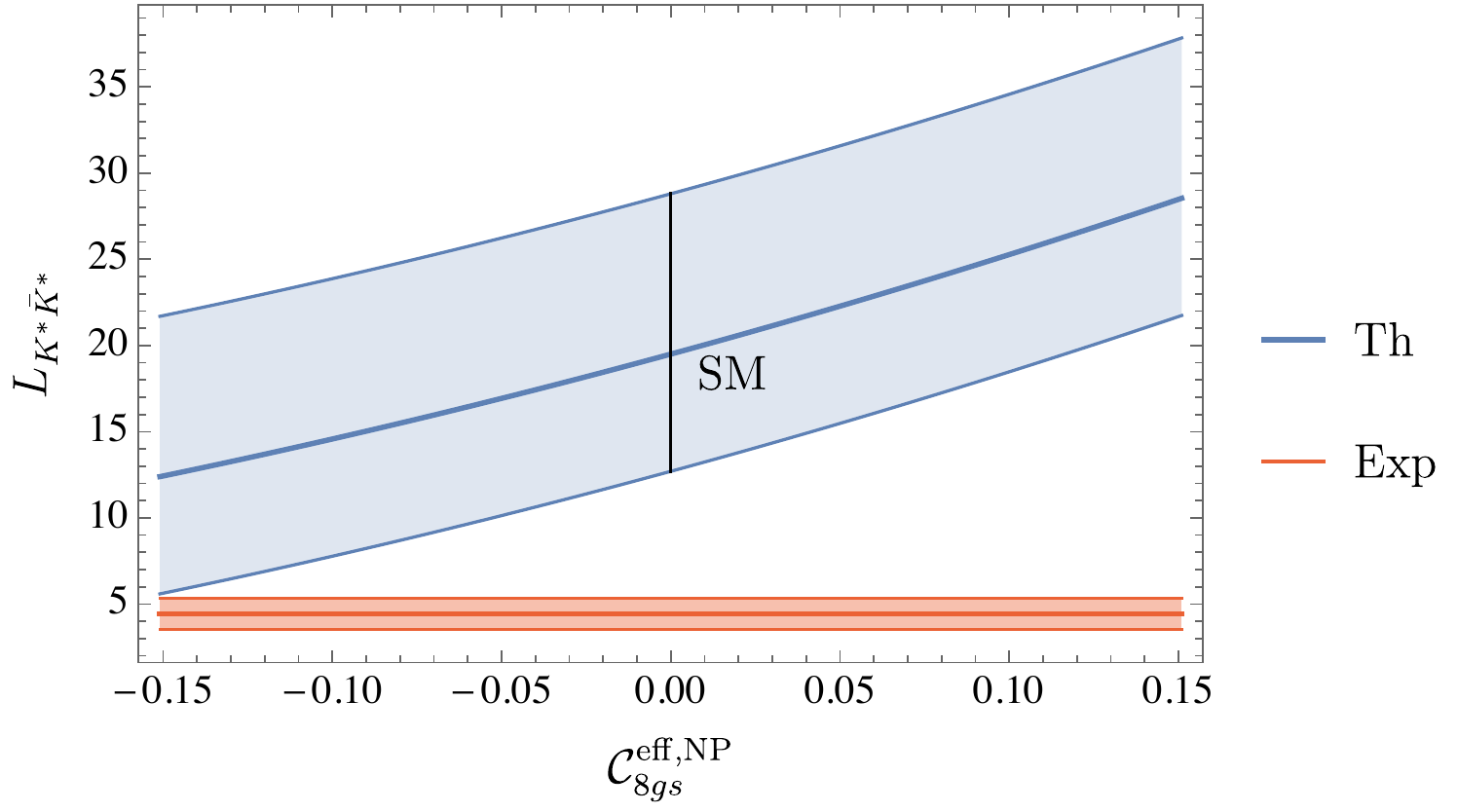}
\caption{\label{fig:C1sC4sC8gs} The tension between the theoretical prediction (blue) and the experimental value (orange) is reduced below $1\sigma$ for ${\cal C}^{\rm NP}_{4s}\simeq 0.25{\cal C}^{\rm SM}_{4s}$ (upper plot) or ${\cal C}^{\rm eff,NP}_{8gs}\simeq -{\cal C}^{\rm eff,SM}_{8gs}$ (lower plot). The predictions are given for ${\cal C}^{\rm NP}_{4s}$ and ${\cal C}^{\rm eff,NP}_{8gs}$  for a range corresponding to 100\% of their respective SM values. }
\end{figure}

In summary, in this article we introduced the $L$-observable as a ratio of longitudinal amplitudes of two non-leptonic related decays governed by a $b \to s$ versus a $b\to d$ transition. This quantity offers the possibility to analyze, in particular, the striking difference in longitudinal polarization fractions between
the two penguin-mediated U-spin partners
$B_{d,s} \to K^{*0} {\bar K^{*0}}$ in a better control way than using the polarization fractions itself. 
We identify the (ratio of) form factors  as the main source
 of theoretical uncertainty for the SM prediction  while 
 WA and HSS infrared divergences entering in a power suppressed way, play a secondary role. We find a tension of 2.6$\sigma$ between our SM prediction and data. Moreover, we identify the main 
  coefficients that can play a role in explaining this anomaly, namely, $C_{4s}$ and $C_{8gs}$ and we discussed a simplified model based on an axi-gluon as a possible explanation (see \cite{non-leptonic} for another example), but requiring a significant amount of  fine-tuning 
  to satisfy $B_s$ mixing.


\begin{figure}
\centering
\includegraphics[width=0.5\textwidth]{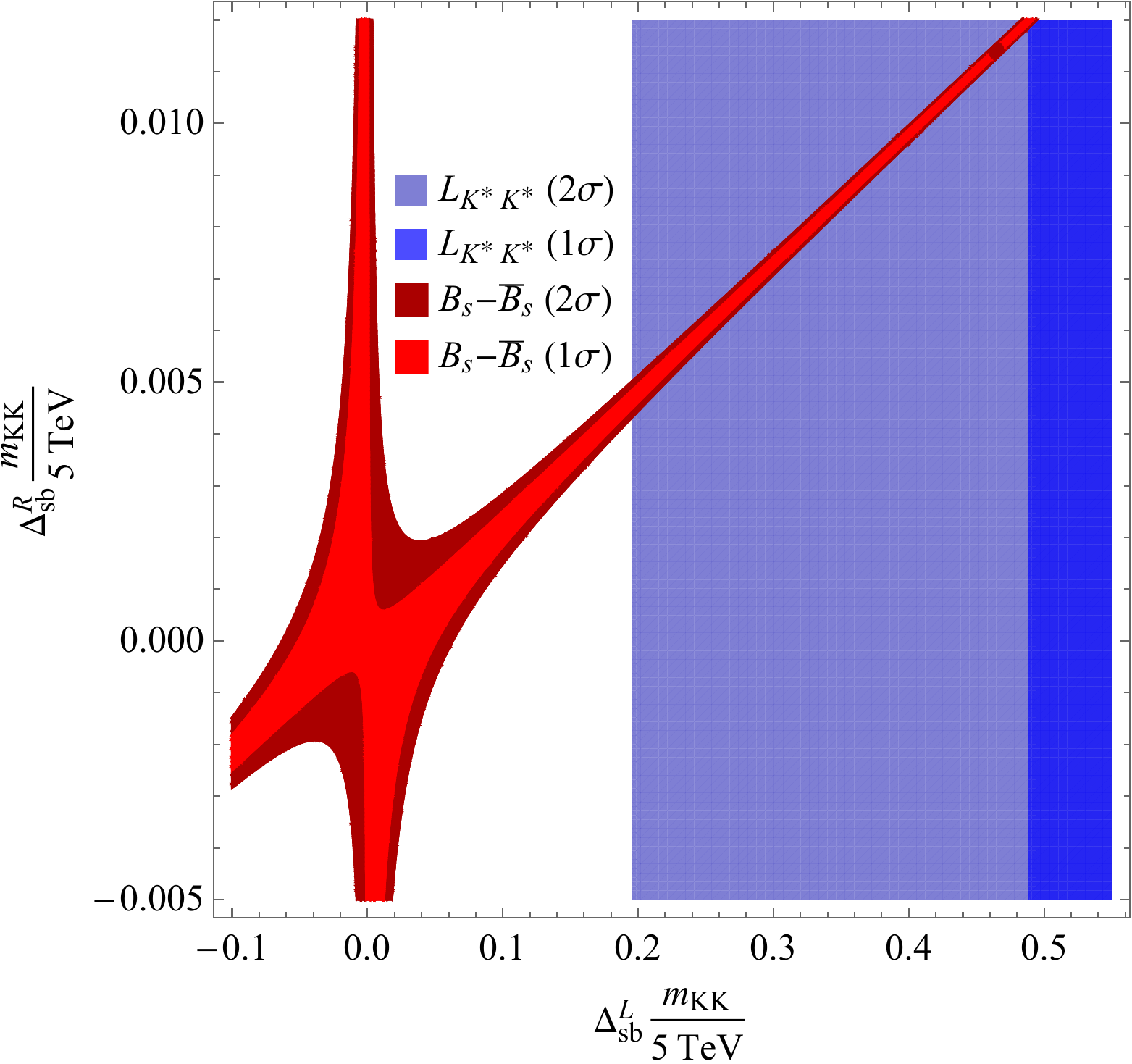}
\caption{\label{KKgluon} 
Preferred regions for the KK-gluon left- and right-handed couplings to s-b quarks from
 $B_s-\bar B_s$ mixing (red) and $L_{K^*\bar{K}^*}$ (blue) compatible with LHC searches assuming real couplings. Note that explaining $L_{K^*\bar{K}^*}$ requires  fine-tuning in $\Delta^L_{sb}$ vs $\Delta^R_{sb}$.}
\end{figure}

\section*{Acknowledgments}
This work received financial support from the Spanish Ministry of Science, Innovation and Universities (FPA2017-86989-P) and the Research Grant Agency of the Government of Catalonia (SGR 1069). JM acknowledges  the financial support by ICREA under the ICREA Academia programme. This project has received support from EU Horizon 2020 programme (Grant No 860881-HIDDeN).

\end{document}